\documentclass[12pt]{article}

\newtheorem{theorem}{Theorem}

\newcommand{\p}{\partial}
\newcommand{\om}{\omega}
\newcommand{\e}{\epsilon}
\renewcommand{\k}{\kappa}

\def\rit{I\!\!R}

\begin{document}

\title{Almost Periodic Passive Tracer Dispersion
\footnote{This work was partly supported by the
USA National Science Foundation Grant DMS-9704345 and
the UK EPSRC Grant GR/M36335.}  }

\author{Hongjun Gao$^{1}$, Jinqiao Duan$^{2}$ and Xinchu Fu$^{3}$
\\
\\
1.  Department of Mathematics\\
 The National University of Singapore, Singapore 119260  \\
 {\em E-mail:  gaohj@math.nus.edu.sg }  \\
\\
2.    Department of Mathematical Sciences \\ 
Clemson University, Clemson, South Carolina 29634, U. S. A.\\ 
  {\em	E-mail:  duan@math.clemson.edu }  
\\
\\	
 \\
3.  Department of Mathematics and Statistics\\
        University of Surrey, Guildford GU2 5XH, U. K.  \\
   {\em   E-mail: mas1xf@ee.surrey.ac.uk}
}


\maketitle

\abstract{The authors investigate the impact of external sources
on the pattern formation of concentration profiles
of passive tracers in a two-dimensional shear flow.
By using the pullback attractor technique for the 
associated nonautonomous
dynamical system,
it is shown that a unique time-almost periodic concentration profile
exists for time-almost periodic external source.
 } 

\bigskip


\newpage

\section{Introduction}

For the benefit of better environment, it is important to understand the
evolution of passive tracers, such as pollutants, temperature or salinity, in
geophysical systems.  Tracers are called passive when they do not dynamically
affect the background fluid velocity field.

The Eulerian approach for studying passive tracer dispersion
attempts to understand the evolution of
tracer concentration profile as a continuous field quantity
\cite{Clark, Schnoor}.

We consider two-dimensional passive tracer dispersion
in a (bounded) shear flow $(u(y), 0)$ such as in a river or in
an oceanic jet.
The passive tracer concentration profile $C(x,y,t)$ then satisfies the
advection-diffusion equation \cite{Clark} 
\begin{eqnarray}
 C_t + u(y)C_x = \k (C_{xx}+C_{yy}) + f(x,y,t),
  \label{eqn}
\end{eqnarray}
where $\k >0$ is the diffusivity
constant, and the source (or sink) term $f(x,y,t)$ accounts for effects
of chemical reactions  \cite{Clark}, external injections
of pollutants  
or heating and cooling  \cite{Schnoor, Hunt}.
The source is generally  dependent on time  or even random in time,
such as random discharge of pollutants into a river or an oceanic jet.

There has been a lot of research on the advection-diffusion
equation {\em without source}; see,
for example, \cite{Clark}, \cite{Smith}, \cite{Young},
 \cite{Mercer} and \cite{Rosencrans}.

In this paper, we study the impact of the external sources
on the  pattern formation   of the concentration profile.
We assume that the  concentration profile satisfies
double-periodic boundary conditions
\begin{eqnarray}
 C, C_x, C_y  \;\; \mbox{are double-periodic in $x$ and $y$ with period $1$},
 \label{BC}
\end{eqnarray}
and appropriate initial condition
\begin{eqnarray}
 C(x,y,0) = C_0(x,y).
 \label{IC}
\end{eqnarray}

Standard abbreviations $\dot{L}_{per}^2$ $=$ $\{u\in L^2(D), u\; \mbox{is} \;D-\mbox{periodic}\; \mbox{and}\; \int_D u = 0\}$, $\dot{H}_{per}^1$ $=$ $\dot{H}_{per}^1(D) = \{u\in H^1(D), u, \nabla u \in \dot{L}_{per}^2\}$,
with $<\cdot, \cdot>$ and $\| \cdot \|$ denoting the  usual scalar product
and norm,
respectively, in $L^2$. We need the following properties and
estimates

{\em  Poincar\'e
inequality\/}
\cite{Gilbarg-Trudinger}
\begin{eqnarray}
\|g\|^2 = \int_D g^2(x,y) \,dxdy \leq  \frac{|D|}{\pi} \int_D |\nabla g|^2
\,dxdy
= \frac{|D|}{\pi} \|\nabla g\|^2
\end{eqnarray}
for $g$ $\in$ $\dot{H}_{per}^1$, and  {\em Young's inequality \/}
\cite{Gilbarg-Trudinger}
\begin{equation}
  AB \leq \frac{\e}2 A^2 + \frac{1}{2\e}B^2,
\end{equation}
where $A, B$ are non-negative real numbers and $\e >0$. \\

In this paper, we prove the following main result.

\begin{theorem} \label{mth}
Assume that $C_0(x, y) \in \dot{L}_{per}^2$, and
$f(x, y, t)$ is temporally almost periodic with its 
$L^2(D)$-norm bounded uniformly in time $t\in \rit$.
Then the model for passive tracer dispersion
(\ref{eqn}) - (\ref{BC})-(\ref{IC}),
has a  unique temporally almost periodic solution that exists for all time $t$
$\in$ $\rit$.
\end{theorem}

\section{Dissipation and Contraction}

In this section we consider the dissipation and contraction properties of
the advection-diffusion
equation with temporally almost periodic source (\ref{eqn}). These 
properties are crucial in the proof of Theorem \ref{mth}
in the next   section.

Integrating   both sides of (\ref{eqn})
with respect to $x, y$ on the domain $D = [0, 1] \times [0, 1]$,
we get

\begin{eqnarray}
&  &  \frac{d}{dt} \int\int  C dxdy   + \int\int u(y)C_x dxdy  \nonumber \\
& = & \k \int\int (C_{xx}+C_{yy}) dxdy + \int\int f(x,y,t)dxdy.
\end{eqnarray}
Note that
$$       \int\int u(y)C_x dxdy = 0      $$
and
$$      \int\int (C_{xx}+C_{yy}) dxdy = 0      $$
due to the double-periodic boundary  conditions (\ref{BC}).
We thus have
\begin{eqnarray}
 \frac{d}{dt} \int\int  C dxdy   = \int\int f(x,y,t)dxdy.
\end{eqnarray}
Here and hereafter, all integrals are with respect to $x, y$
over $D$.
Thus, when there is no source, the spatial average or mean of the
concentration $C(x,y,t)$ does not change with time.
When there is a source, the time-evolution of
the spatial average of    $C(x,y,t)$
is determined only by the source term.
In order to understand more delicate
impact of source on the evolution  of
$C(x,y,t)$, it is appropriate to assume that
the source has zero spatial average or mean:
\begin{eqnarray}
\int\int f(x,y,t)dxdy = 0.
\end{eqnarray}
With such a source, the mean of $C(x,y,t)$
is a constant. Without loss of generality or
after removing the non-zero constant by a translation,
we may assume that $C(x,y,t)$ has zero-mean.
So we study the dynamical behavior of
$C(x,y,t)$ in zero-mean spaces.

Note that the linear operator $-\k (\p_{xx}+\p_{yy}) +u(y)\p_x$
is sectorial (\cite{Henry}, p. 19) in $\dot L^2_{per} (D)$. Thus
if $f(x,y,t)$ has continuous derivative in time $t$,
the linear system (\ref{eqn}),
(\ref{BC}), (\ref{IC}) has a unique strong solution
for every $C_0(x,y)$ in $\dot L^2_{per} (D)$
(\cite{Henry}, p. 52).

Define the solution operator $S_{t,t_0}$ $:$ $L^2$ $\to$ $L^2$  by
$S_{t,t_0} \om_0$ $:=$ $\om(t)$ 
for $t$ $\geq$ $t_0$, where $\om(t)$ is the
solution of the QG equations in $L^2$ starting 
at $\om_0 $ $\in$ $L^2$
at time $t_0$.  Since the the dissipative   system
(\ref{eqn})-(\ref{BC})-(\ref{IC}) are
strictly parabolic, the solution operators $S_{t,t_0}$ exist and are
compact for all $t$ $>$ $t_0$; 
see, for example, \cite{Henry}.  In fact, the $S_{t,t_0}$
are compact in
$H^k_0$ for all $k \geq 0$ and so, in particular, $S_{t,t_0}B$ is a compact
subset
of  $L^2$  for each $t$ $>$ $t_0$ and every closed and bounded subset $B$
of $L^2$.\\

We now show that this system  is a dissipative system in the sense
    \cite{Hale}  that
all solutions $C(x, y, t)$ approach a bounded set in $\dot L^2_{per} (D)$ as
time goes to infinity.
Multipling (\ref{eqn}) by $C(x, y, t)$ and
integrating over $D$, we get

\begin{eqnarray}
&   & \frac12 \frac{d}{dt}\|C\|^2  + \int\int u(y)C_x C dxdy  \nonumber  \\
& = &  - \k \int\int | \nabla C |^2 dxdy + \int\int f(x,y,t)C dxdy.
        \label{estimate1}
\end{eqnarray}

Note that, using the double-periodic boundary  conditions (\ref{BC}),
\begin{eqnarray}
   \int\int u(y)C_x C dxdy = 0 .
        \label{estimate2}
\end{eqnarray}

We further assume that the square-integral of $f(x,y, t)$ with respect to
$x, y$ is bounded in time, i.e. $\|f\| \le M$($ M > 0 $ is a constant independent of $t$).
Then, by  the Young inequality,
\begin{eqnarray}
 \int\int f(x,y,t)C dxdy  & \leq &  \frac1{2\e} \int\int |f(x,y,t)|^2 dxdy
                        + \frac{\e}2 \int\int |C|^2 dxdy     \nonumber  \\
        & \leq & \frac{M}{2\e}  + \frac{\e}2 \int\int |C|^2 dxdy,
                \label{estimate3}
\end{eqnarray}
where 
$\e>0$ is an arbitrary positive number.

Since $C$ has zero mean, we can use the Poincar\'e  inequality
(\cite{Gilbarg-Trudinger}, p. 164) to obtain
\begin{eqnarray}
  \|C\|^2   \leq    \frac{2}{\pi}  \| \nabla C \|^2.
      \label{estimate4}
\end{eqnarray}

Putting $(\ref{estimate2}), (\ref{estimate3}), (\ref{estimate4})$
into $(\ref{estimate1})$, we obtain
\begin{eqnarray}
\frac12 \frac{d}{dt}\|C\|^2
& \leq &  (\frac{\e}2 - \frac{\k\pi}{2}) \| C \|^2 + \frac{M}{2\e},
        \label{estimate5}
\end{eqnarray}
or
\begin{eqnarray}
 \frac{d}{dt}\| C \|^2
& \leq  &  ( \e - \pi\k) \| C \|^2 + \frac{M}{\e}.
\end{eqnarray}
We now fix $\e>0$ so small that $\e - \frac{\k}{\pi}<0$.
By the Gronwall inequality (\cite{Henry}), we finally get

\begin{eqnarray}
 \|C\|^2  \leq (\|C_0\|^2  +
        \frac{M}{\e( \e - \frac{\k}{\pi})} )e^{( \e - \frac{\k}{\pi})t}
       +  \frac{M}{\e(\frac{\k}{\pi}-\e)}.
\end{eqnarray}
Hence all solutions $C(x,y,t)$ enter a bounded
set in $\dot L^2_{per}$,
$$
{\cal B} = \{ C:  \;\;   \|C\| \leq  \sqrt{ \frac{M}{\e(\frac{\k}{\pi}-\e)} } \},
$$
as time goes to infinity. The system (\ref{eqn}) is  therefore a
dissipative system.

We now consider the strong contraction property.
Assume that  $C^{(i)}$  are two trajectories    
corresponding to initial values
$C^{(i)}_0$ $\in$ ${\cal B}$, $i$ $=$ $1$ and $2$. Note that these
trajectories remain inside ${\cal B}$. Their difference $\delta C$ $=$
$C^{(1)} - C^{(2)}$ satisfies the equation
$$
\delta C_t + u(y)\delta C_x = \k(\delta C_{xx} + \delta C_{yy}).$$
Similarly to the proof above it can be shown from this equation that
\begin{equation}
\frac{1}{2}\frac{d}{dt}\|\delta C\|^2 + \int_D u(y)\delta C_x\delta C dxdy
= - \k \|\nabla \delta C\|^2.\label{diff}
\end{equation}
By (4) and $\int_D u(y)\delta C_x\delta C dxdy = 0$, (\ref{diff}) can be written as 
$$
\frac{1}{2}\frac{d}{dt}\|\delta C\|^2 + \k\pi\|\delta C\|^2 \le 0.
$$
This gives 
$$\|\delta C\|^2 \le \|\delta C_0\|^2e^{-2\k\pi t}\to 0, \;\mbox{as}\; t\to \infty.$$
This is the   strong contraction condition.

\section{Almost periodic dynamics}

A function $\varphi$  $:$ $\rit$ $\to$ $X$, where $(X,d_X)$ is  a
metric space, is called {\em almost periodic\/} \cite{Besicovitch}
 if for every $\varepsilon$ $>$ $0$  there exists a
relatively dense subset $M_{\varepsilon}$ of $\rit$ such that
$$
d_X \left(\varphi (t+ \tau ), \varphi (t) \right) <  \varepsilon
$$
for all $t$ $\in$ $\rit$ and $\tau$ $\in M_{\varepsilon }$.  A subset $M$
$\subseteq$ $\rit$ is called {\em relatively dense} in $\rit$ if there exists
a positive number $l$ $\in$ $\rit$ such that for every $a$ $\in$ $\rit$ the
interval $[a,a+l]\bigcap \rit$ of length $l$ contains an element of $M$, i.e.
$M\bigcap [a,a+l]$ $\ne$ $\emptyset$ for every $a$ $\in$ $\rit$.

In order to study   the temporally almost periodic solutions
for the passive tracer convection-diffusion equation (\ref{eqn}),
we need some results from the theory of nonautonomous dynamical systems.
Consider first an autonomous dynamical system on a metric space $P$
described by
a group $\theta$ $=$ $\{\theta_t\}_{t \in \rit}$ of mappings of $P$ into
itself.

Let $X$ be a complete metric space and consider a continuous mapping
$$
 \Phi : \rit^{+} \times P \times X \to   X
$$
satisfying the properties
$$
\Phi(0,p,\cdot) = {\rm id}_X, \qquad
\Phi(\tau +t,p,x)  =  \Phi(\tau,\theta_t  p, \Phi(t,p,x))
$$
for all $t$, $\tau$ $\in$ $\rit^{+}$, $p$ $\in$ $P$ and $x$ $\in$ $X$.
The mapping $\Phi$ is called a cocycle on $X$ with respect to $\theta$ on $P$.

The appropriate concept of an attractor for a nonautonomous cocycle systems is
the {\em pullback attractor\/}. In contrast to autonomous attractors it
consists of a family subsets of the original state space $X$ that are
indexed by
the cocycle parameter set.

 A family $\widehat{A}$ $=$ $\{A_p\}_{p \in P}$ of nonempty compact sets of
$X$ is
called a {\rm pullback attractor\/} of the cocycle $\Phi$ on $X$ with
respect to
$\theta_t$ on $P$ if it is ${\Phi}$--invariant, i.e.
$$
\Phi(t,p,A_p) = A_{\theta_t} p  \qquad \mbox{for all} \quad t \in \rit^{+},
p \in P,
$$
and {\rm pullback attracting}, i.e.
$$
\lim_{t \to \infty} H^{*}_X\left(\Phi(t,\theta_{-t}p,D), A_p\right) = 0
\qquad \mbox{for all} \quad D \in K(X), \  p \in P,
$$
where $K(X)$ is the space of all nonempty compact subsets of the metric
space $(X,d_X)$.
 Here $H^{*}_X$ is the Hausdorff 
semi--metric between nonempty compact subsets
of $X$, i.e. $H^{*}_X(A,B)$ $:=$ $\max_{a \in A} {\rm dist}(a,B)$ $=$
$\max_{a\in A} \min_{b\in B} d_X(A,b)$ for $A$, $B$ $\in$ $K(X)$.

The following theorem combines  several known results.  See Crauel and
Flandoli
\cite{CF}, Flandoli and Schmalfu{\ss} \cite{FS1}, and Cheban \cite{C1} as
well as
\cite{CKS,KS2}  for this and various related proofs.

\begin{theorem} \label{th1}
Let $\Phi$ be a continuous cocycle on a metric space $X$ with respect to a
group
$\theta$ of continuous mappings on a metric space $P$. In addition, suppose
that
there is a nonempty compact subset $B$ of $X$ and that for 
every $D$ $\in$
 $K(X)$
there exists a $T(D)$ $\in$ $\rit^{+}$, which is independent of $p$ $\in$ $P$,
 such that
\begin{equation}\label{fa}
\Phi(t,p,D) \subset B \quad \mbox{for all} \quad t > T(D).
\end{equation}
Then  there exists a unique pullback attractor
$\widehat{A}$ $=$
$\{A_p\}_{p \in P}$ of the cocycle $\Phi$ on $X$, where
\begin{equation}\label{pbat}
A_p = \bigcap_{\tau \in \rit^{+}} \overline{\bigcup_{t > \tau \atop t \in
\rit^{+}}
\Phi\left(t,\theta_{-t}p,B\right)}.
\end{equation}
Moreover, if the cocycle $\Phi$  is strongly
contracting inside
the absorbing set $B$. Then the pullback attractor consists of  singleton
valued
sets, 
i.e.   $A_p$ $=$ $\{a^*(p)\}$,  and the 
mapping $p$ $\mapsto$
$a^*(p)$ is
continuous.
\end{theorem}

The  solution operators $S_{t,t_0}$ for (1) form a cocycle 
mapping on $X$ $=$
$\dot{L}_{per}^2$ with
parameter set $P$ $=$ $\rit$, where $p$ $=$ $t_0$, the initial time, and
$\theta_t t_0$ $=$ $t_0+t$, the left shift by time $t$. However, the space
$P$ $=$
$\rit$ is not compact here. Though more complicated, it is more useful to
consider
$P$ to be the closure of the subset $\{\theta_t f, t \in \rit\}$, i.e. the
hull of $f$,  in
the metric  space $L^2_{loc}\left(\rit,\dot{L}_{per}^2(D)\right)$ of locally
$L^2(\rit)$--functions
$f$ $:$ $\rit$ $\to$ $\dot{L}_{per}^2(D)$ with the metric
$$
d_P(f,g) :=  \sum_{N=1}^{\infty} 2^{-N} \min\left\{1, \sqrt{\int_{-N}^N
\|f(t)-g(t)\|^2 \, dt} \right\}
$$
with  $\theta_t$ defined to be  the left shift operator, i.e. 
$\theta_t f(\cdot) := f(\cdot+t)$.  
By a classical result \cite{Besicovitch,Sell},  a function
$f$ in the
above metric space is almost periodic if and only if the the  hull of $f$
is compact
and minimal. Here minimal means nonempty, closed and invariant with respect
to the
autonomous dynamical system generated by the shift operators $\theta_t$
such that
with no proper subset has these properties. The cocycle mapping is defined
to be the
solution $C(t)$ of (1) starting at $C_0$ at time $t_0$ $=$ $0$ for a given
forcing
mapping $f$ $\in$ $P$, i.e.
$$
\Phi(t,f,\om_0) := S_{t,0}^f \ \om_0,
$$
where we have included a superscript $f$ on $S$ to denote the dependence on
the forcing
term $f$. (This dependence is in fact continuous). The cocycle property
here follows
from the fact  that 
$S_{t,t_0}^f \om_0$ $=$ $S_{t-t_0,0}^{\theta_{t_0}f}\
\om_0$ for
all $t$ $\geq$ $t_0$, $t_0$ $\in$ $\rit$, 
$C_0$ $\in$ $\dot{L}_{per}^2$ 
and $f$ $\in$
$P$. 

Following Theorem 
2 
and 
the dissipativity and  
contractivity results which we have obtained in the last section,
we conclude that   the passive tracer convection-diffusion 
model (\ref{eqn})-(\ref{BC})-(\ref{IC}) has 
the unique pullback attractor,
  consists of the  singleton valued component  
  $\{a^*(p)\}$ and the mapping
$p$ $\mapsto$ $a^{*}(p)$ is continuous on $P$. 
As in Duan and Kloeden \cite{Duan-Kloeden}, 
we now show that this
singleton attractor $a^{*}(p)$ defines an almost periodic 
solution.

In fact, the mapping  $p$ $\mapsto$ $a^{*}(p)$ 
is uniformly continuous on $P$ because $P$ is compact subset of
$L^2_{loc}\left(\rit,L^2(D)\right)$ due to the assumed  almost periodicity.
That is, for every $\varepsilon$ $>$ $0$ there exists a $\delta(\varepsilon)$
$>$ $0$ such that $\|a^*(p)-a^*(q) \|$ $<$ $\varepsilon$ whenever
$d_P(p,q)$ $<$
$\delta$.  Now let  the point $\bar{p}$ ($=$ $f$, the given temporal
forcing function)
be almost periodic and   for $\delta$ $=$ $\delta(\varepsilon)$
$>$ $0$ denote by $M_{\delta}$ the relatively  
dense subset of $\rit$ such that
$d_P(\theta_{t+\tau}\bar{p},\theta_t \bar{p})$ $<$ $\delta$ 
for all $\tau$ $\in$ $M_{\delta}$ and 
$t$ $\in$ $\rit$. From this and the uniform
continuity we have
$$
\|a^*(\theta_{t+\tau} \bar{p}) - a^*(\theta_t \bar{p})\| < \varepsilon
$$
for all $t$ $\in$ $\rit$ and $\tau$ $\in$ $M_{\delta(\varepsilon)}$. Hence
$t$ $\mapsto$ $C^*(t)$ $:=$  $a^*(\theta_{t}\bar{p})$ is almost periodic, and 
it is a solution of the   passive tracer convection-diffusion model.
It is unique as the single-trajectory pullback attractor is the only trajectory
that exists and is bounded for the entire time line.
Therefore, the conclusion in Theorem 1 follows.
 
\bigskip 
 
  {\bf Acknowledgement.}
  We thank Peter E. Kloeden (Frankfurt, Germany)
  for introducing 
  us to the study of
    almost periodicity in nonautonomous dynamical systems.

\end{document}